\documentstyle[lprocl,colordvi,epsfig,11pt]{article}
\textheight
8.5in  \parskip 7pt \openup1\jot %\parindent=0.5in
\def\st{\scriptstyle}

\def\be{\begin{equation}}
\def\ee{\end{equation}}
\def\bea{\begin{eqnarray}}
\def\eea{\end{eqnarray}}
\newskip\humongous \humongous=0pt plus 1000pt minus 1000pt
\def\caja{\mathsurround=0pt}
\def\eqalign#1{\,\vcenter{\openup1\jot \caja
        \ialign{\strut \hfil$\displaystyle{##}$&$
        \displaystyle{{}##}$\hfil\crcr#1\crcr}}\,}
\newif\ifdtup

\def\eqright #1\cr{\noalign{\hfill$\displaystyle{{}#1}$}}
\def\eqleft #1\cr{\noalign{\noindent$\displaystyle{{}#1}$\hfill}}
\def\oldreffmt#1{\rlap{[#1]} \hbox to 2\parindent{}}

\def\figfmt#1{\rlap{Figure {#1}} \hbox to 1in{}}

\def\begineq #1\endeq{$$ \refstepcounter{equation}\eqalign{#1}\eqno
	(\theequation) $$}
\def\contlimit{\,{\hbox{$\longrightarrow$}\kern-1.8em\lower1ex
\hbox{${\scriptstyle (a\rightarrow0)}$}}\,}
\def\centeron#1#2{{\setbox0=\hbox{#1}\setbox1=\hbox{#2}\ifdim
\wd1>\wd0\kern.5\wd1\kern-.5\wd0\fi
\copy0\kern-.5\wd0\kern-.5\wd1\copy1\ifdim\wd0>\wd1
\kern.5\wd0\kern-.5\wd1\fi}}
\def\centerover#1#2{\centeron{#1}{\setbox0=\hbox{#1}\setbox
1=\hbox{#2}\raise\ht0\hbox{\raise\dp1\hbox{\copy1}}}}
\def\centerunder#1#2{\centeron{#1}{\setbox0=\hbox{#1}\setbox
1=\hbox{#2}\lower\dp0\hbox{\lower\ht1\hbox{\copy1}}}}
\def\lsim{\;\centeron{\raise.35ex\hbox{$<$}}{\lower.65ex\hbox
{$\sim$}}\;}
\def\gsim{\;\centeron{\raise.35ex\hbox{$>$}}{\lower.65ex\hbox
{$\sim$}}\;}
\def\st#1{\centeron{$#1$}{$/$}}
\def\super#1{\ifmmode \hbox{\textsuper{#1}}\else\textsuper{#1}\fi}
\def\textsuper#1{\newcount\holdspacefactor\holdspacefactor=\spacefactor
$^{#1}$\spacefactor=\holdspacefactor}
\def\getcite#1,{\advance\citenumber by1
\ifnum\citenumber=1
\ref{#1}\let\next=\getcite\else\ifx#1@\let\next=\relax
\else ,\ref{#1}\let\next=\getcite\fi\fi\next}
\def\upon #1/#2 {{\textstyle{#1\over #2}}}
\relax

\def\til#1{\centeron{\hbox{$#1$}}{\lower 2ex\hbox{$\char'176$}}}
\def\tild#1{\centeron{\hbox{$\,#1$}}{\lower 2.5ex\hbox{$\char'176$}}}
\def\sumtil{\centeron{\hbox{$\displaystyle\sum$}}{\lower
-1.5ex\hbox{$\widetilde{\phantom{xx}}$}}}

\def\kbar{\underline{k}}

\def\pom{{\rm P\kern -0.53em\llap I\,}}
\def\spom{{\rm P\kern -0.36em\llap \small I\,}}
\def\sspom{{\rm P\kern -0.33em\llap \footnotesize I\,}}

\begin{document} 

\begin{titlepage} 

\rightline{\vbox{\halign{&#\hfil\cr
&ANL-HEP-CP-98-31 \cr
&\today\cr}}} 
\vspace{1.25in} 

\begin{center}
{\bf THE SUPERCRITICAL POMERON IN QCD}\footnote{Work 
supported by the U.S.
Department of Energy, Division of High Energy Physics, \newline Contracts
W-31-109-ENG-38 and DEFG05-86-ER-40272} 
\medskip

Alan. R. White\footnote{arw@hep.anl.gov }
\end{center}
\vskip 0.6cm

\centerline{High Energy Physics Division}
\centerline{Argonne National Laboratory}
\centerline{9700 South Cass, Il 60439, USA.}
\vspace{0.5cm}

\begin{abstract} 

Deep-inelastic diffractive scaling violations have provided fundamental
insight into the QCD pomeron, suggesting a single gluon inner structure rather
than that of a perturbative two-gluon bound state. This talk outlines a 
derivation of a high-energy, transverse momentum cut-off, confining solution
of QCD. The pomeron, in first approximation, is a single reggeized gluon
plus a ``wee parton'' component that compensates for the color and particle
properties of the gluon. This solution corresponds to a supercritical phase
of Reggeon Field Theory.  

\end{abstract} 

\vspace{2in}
\begin{center}

Invited Talk Presented at the LAFEX
International Workshop on Diffractive Physics,
\newline Rio de Janeiro, Brazil, February 16-20, 1998.

\end{center}

\end{titlepage}

\section{Introduction}

A complete understanding of the pomeron in QCD requires the solution of the
theory at high-energy. Although high-energy, perhaps, suggests a 
perturbative starting point, essential physics is clearly absent in
the QCD perturbation expansion. In particular, it is well-established that 
confinement and chiral symmetry breaking are ``low energy'' properties that 
are essential in the physical solution of QCD but are not present in 
perturbation theory. We will also focus on two ``high energy'' experimental
properties of the pomeron which are not present in perturbation theory. In 
small momentum transfer processes the pomeron is (approximately) a Regge 
pole, while in large $Q^2$ deep-inelastic scattering it looks remarkably like 
a single gluon~\cite{h1}. In this talk I will outline a high-energy solution
of QCD in which these high energy properties of the pomeron are closely
related to the low-energy ``non-perturbative'' properties of confinement and
chiral symmetry breaking. 

It is eighteen years since I first proposed~\cite{arw80} identifying a 
``supercritical phase'' of Reggeon Field Theory (RFT) with a 
semi-perturbative picture of the QCD pomeron. The suggestion was
that, in the supercritical  phase, the ``pomeron is a single (reggeized)
gluon in a soft gluon background''. After many tries, I have finally found a
detailed derivation~\cite{arw97} of this connection between RFT and the QCD
pomeron. In this talk I will outline the derivation and show how the
desired ``non-perturbative'' properties emerge. 

In the course of this work I have gradually realized that many 
fundamental properties of the physical solution of QCD are deeply 
inter-related with the nature of the pomeron. As I will briefly 
elaborate at the end of this talk, I now believe that my 
pomeron solution also solves, in principle, the problem of finding 
a light-cone wee parton distribution for physical states which reproduces
all the properties normally associated with a non-perturbative vacuum. Such
a distribution is believed by many~\cite{kw} to be behind the success of the
constituent quark model in describing low-energy QCD. It also potentially
provides a deeper origin of the parton model in QCD than that provided by
the factorization properties of leading twist perturbation theory. 

Multi-regge theory provides the framework for my analysis. By using 
reggeon unitarity equations~\cite{gpt,arw1} well-known Regge limit QCD
calculations~\cite{bfkl,bs,fs1} can be extended to obtain multiparticle
amplitudes involving multiple exchanges of reggeized gluons
and quarks in a variety of Regge channels. In particular we can 
study amplitudes in which reggeon bound states and their scattering 
amplitudes appear. (Presently this is impossible in any other formalism.).  In 
Fig.~1 we show qualitatively how we expect pion Regge pole amplitudes, for
scattering via pomeron exchange, to emerge from the 
multiparticle reggeon diagrams describing the scattering of multiquark 
states. We will find that new ``reggeon helicity-flip'' vertices that I have 
calculated play a vital dynamical role in such amplitudes. 

In fact the hadron amplitudes we obtain are initially selected by a
(``volume'') infra-red divergence that appears when SU(3) gauge symmetry is
partially broken to SU(2) and the 
limit of zero quark mass is also taken. The divergence is
produced by quark loop helicity-flip 
\newpage 
~
\vspace{0.3in}

\noindent \parbox{4in}{
\leavevmode
\epsfxsize=3.8in
\epsffile{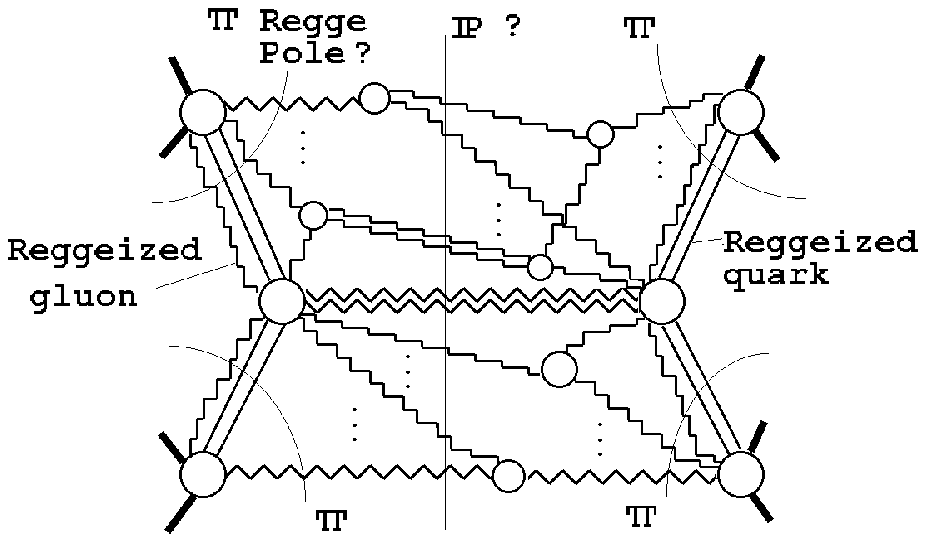}
}
\parbox{2in}{
\leavevmode
\epsfxsize=1.8in
\epsffile{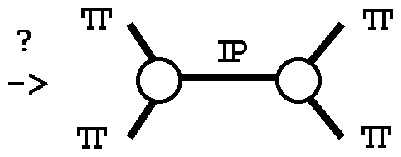}
}
\begin{center}
Fig.~1 The Anticipated Formation of Pion 
Scattering Amplitudes
\end{center}

\noindent vertices involving chirality
violation (c.f. instanton interactions). 
The chirality violation survives the massless quark limit 
because of an infra-red effect of the triangle anomaly~\cite{cg}. 
We show that the divergence
produces what we will call a ``wee parton condensate'' 
which is directly responsible (when the gauge symmetry is
partially broken) for confinement and chiral symmetry breaking. The pomeron is 
(in first approximation) a reggeized gluon in the wee
parton condensate and so is obviously a Regge pole. Although we will not give 
any description of supercritical RFT~\cite{arw1} in this talk we do find
that all the essential features are present. We briefly discuss the 
restoration of SU(3) gauge symmetry. It is closely related with the critical 
behaviour of the pomeron~\cite{cri}
and the consequent disappearance of the supercritical condensate. We note 
that the large $Q^2$ of deep-inelastic scattering provides a 
finite volume constraint that can keep the theory (locally) in the
supercritical phase as the full gauge symmetry is restored. A single gluon
(in the background wee parton condensate) should then be a good
approximation for the pomeron.

\section{Multi-Regge Theory} 

This is an abstract formalism based on the existence of asymptotic
analyticity domains for multiparticle amplitudes derived~\cite{arw1,sw}
via ``Axiomatic
Field Theory'' and ``Axiomatic S-Matrix Theory''. All the assumptions made
are expected to be valid in a completely massive spontaneously-broken gauge
theory. Since we begin with massive reggeizing gluons, this is
effectively the starting point for our analysis of QCD. We can very briefly
list the key ingredients as follows. 

\noindent {\it i)  Angular Variables } 
\newline For an N-point amplitude we can introduce 
variables corresponding to any Toller diagram, i.e. any tree diagram, drawn as 
in Fig.~2, that involves
only three-point vertices. The result 
\newline \parbox{2.9in}{ \openup\jot is that we can write 
$$
M_N(P_1,..,P_N) \equiv
M_N\left(t_1,..,t_{N-3},g_1,..,g_{N-3}\right)
$$
where $t_j=Q_j^2$ and $g_j $ is in the little group of $Q_j$, i.e. 
for $t_j > 0$, $g_j \in$ SO(3), and for $t_j < 0$, 
$g_j \in$ SO(2,1). A set of 3N - 10 independent 
variables is obtained, 
N-3 $~t_i$ variables, 
N-3 $~z_j ~(\equiv \cos\theta_j)$ variables 
and 
N-4 $~u_{jk}~ (\equiv e^{i(\mu_j - \nu_k)})$ variables.

}
\parbox{3.1in}{
\begin{center}
\leavevmode
\epsfxsize=2.5in
\epsffile{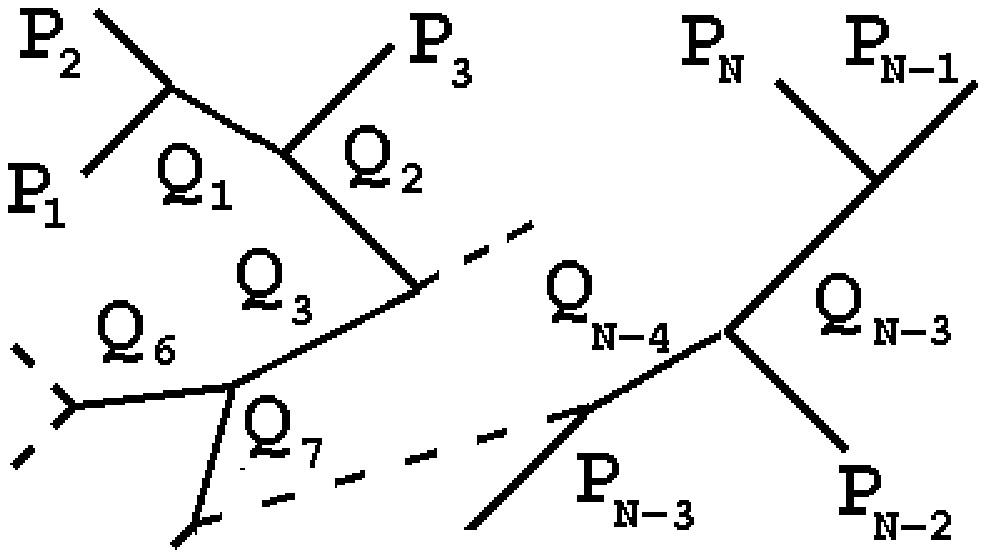}
$~$ \newline 
Fig.~2 A Tree Diagram with Three Point Vertices.
\end{center}
}

\noindent ii) {\it Multi-Regge Limits } 
\newline These limits are defined by  
$z_j \to \infty ~,~\forall j$. We will also be interested in 
{\it Helicity-Pole Limits} in which some $u_{jk} 
\to \infty$ and some $z_j \to \infty$. In a helicity-pole limit a smaller 
number of invariants is taken large. 

\noindent iii) {\it Partial-wave Expansions}
\newline Using 
$ f(g)=\sum^\infty_{J=0}\,\sum_{|n|,|n'|<J}D^J_{nn'}(g)a_{J nn'}$, 
for a function $f(g)$ defined 
on SO(3), leads to 
$$
M_N(\til{t},\til{g})=\sum_{\til{J},\til{n},\til{n'}}
\prod_i~D^{J_i}_{n_in_i'}(g_i)~ a_{\til{J},\til{n},\til{n'}} (\til{t})
$$

\noindent iv) {\it Asymptotic Dispersion Relations}
\newline We can write $~~~~M_N~=~\sum_{C}M_N^C +M^0~~~~$ where
$$
M_N^C ~=~ {1\over (2\pi i)^{N-3}}
\int \frac{ ds'_1\ldots ds'_{N-3}\Delta^C(
..t_i.,..u_{jk}.,..s'_i.)}
{(s'_1-s_1)(s'_2-s_2)\ldots (s'_{N-3}-s_{N-3})} 
$$
and $\sum_{C}$ is over all sets of (N-3) Regge limit asymptotic cuts.
$M^0$ is non-leading in the multi-regge limit. The resulting separation into
(hexagraph) spectral components is crucial for the development of
multiparticle complex angular momentum theory. 

\noindent v) {\it Sommerfeld-Watson Representations 
of Spectral Components} 
\newline A multiple transformation of the partial-wave expansion gives e.g.
$$
\eqalign{ M^C_4=&{1\over 8}\sum_{{\scriptstyle N_1, N_2}} \int
{dn_2  dn_1 dJ_1 ~u_2^{n_2} u_1^{n_1}
d^{J_1}_{0,n_1}(z_1)
d^{n_1+N_2}_{n_1,n_2}(z_2)d^{n_2+N_3}_{n_2,0}(z_3)
\over
\sin\pi n_2\sin\pi(n_1-n_2)\sin\pi(J_1-n_1)}~a^C_{N_2N_3}(J_1,n_1,n_2,
\til{t})\cr
& ~+~~\sumtil_{\til{\scriptstyle J}\til{\scriptstyle
n}}d^{J_1}_{0,n_1}(z_1)u_1^{n_1}d^{J_2}_{n_1,n_2}
(z_2)u_2^{n_2}d^{J_3}_{n_2,0}(z_3)a_{\til{\scriptstyle J}\til{\scriptstyle
n}}(\til{t})
}
$$
These representations give the form of the asymptotic behaviour in both
multi-Regge and helicity-pole limits. In particular, 
in a ``maximal'' helicity-pole limit, 
in which the maximal number of $u_{jk} 
\to \infty$, only a single (analytically-continued) partial-wave amplitude 
appears.

\noindent vi) {\it $t$-channel Unitarity in the $J$-plane}
\newline Multiparticle unitarity in every $t$-channel can be partial-wave
projected, diagonalized, and continued to complex $J$ as an equation for 
partial-wave amplitudes, i.e. 
$$ 
a^+_J - a^-_J= i\int d\rho \sum_{\til{N}} 
\int {dn_1 dn_2  \over 
sin\pi(J-n_1-n_2) }\int {dn_3 dn_4 \over sin\pi(n_1 -n_3 -n_4)} ~\cdots 
~a^+_{J\til{N}
\til{n}}a^-_{J\til{N}\til{n}} 
$$
Regge poles at $n_i=\alpha_i$, together with the phase-space 
$\int d\rho $ and the ``nonsense poles'' at 
$J= n_1 +n_2 -1, n_1=n_3 + n_4 -1 , ~...~$ generate multi-reggeon 
thresholds, i.e. Regge cuts. 

\noindent vii) {\it Reggeon Unitarity }
\newline In ANY $J$-plane of any partial-wave amplitude, the ``threshold''
discontinuity due to $M$ Regge poles with trajectories 
$\til{\alpha}= (\alpha_1, \alpha_2, \cdots \alpha_M)$
is given by the reggeon unitarity equation 
$$ 
\centerunder{disc}{\raisebox{-3mm}{$\scriptstyle J=\alpha_M(t)$}}~~ 
a_{\til{N} \til{n}}(J) 
~=~ {\xi}_{M} \int d\hat\rho~
a_{\til{\alpha}}(J^+)
a_{\til{\alpha}}(J^-)
{\delta\left(J-1-\sum^M_{k=1}
(\alpha_k-1)\right)\over \sin{\pi\over 2}(\alpha_1-\tau'_1)\ldots\sin{\pi\over
2}(\alpha_M-\tau'_M) }
$$
Writing $t_i=k_i^2~~$ (with 
$\int dt_1 dt_2 \lambda^{-1/2}(t,t_1,t_2) =2\int d^2 k_1 d^2 k_2 \delta^2(k 
- k_1 - k_2) $), $\int d \hat{\rho} $ can be written in terms of
two dimensional ``$~k_{\perp}$'' integrations, 
anticipating the reggeon diagram results of 
direct $s$-channel high-energy calculations~\cite{bfkl,bs,fs1}. 
The generality of reggeon unitarity makes it particularly powerful when 
applied to the
partial-wave amplitudes appearing in maximal helicity-pole limits. 

\section{Reggeon Diagrams in QCD}

Leading-log Regge limit calculations of elastic and multi-regge production 
amplitudes in (spontaneously-broken) gauge theories
show\cite{bfkl,bs,fs1} that both gluons and quarks ``reggeize'', i.e.
they lie on Regge trajectories. Non-leading log calculations are described
by ``reggeon diagrams'' involving reggeized gluons and
quarks. In fact, reggeon unitarity
requires that higher-order calculations produce a complete set of reggeon
diagrams. 

Gluon reggeon diagrams involve a reggeon propagator for each reggeon state 
and also gluon particle poles e.g. 
$$
\hbox{two-reggeon state} ~~~\leftrightarrow ~~
\int {d^2k_1 \over (k_1^2 +M^2) } {d^2k_2 \over (k_2^2 + M^2)}~~ 
{\delta^2(k_1'+k_2'-k_1-k_2)
\over J-1 +  \Delta(k_1^2,M^2) + \Delta(k_2^2,M^2)}
$$
The BFKL equation\cite{bfkl}
corresponds to 2-reggeon unitarity,
i.e. iteration of the color-zero 2-reggeon state with the 2-2 reggeon
interaction 
$$                        
\Gamma_{22}(\kbar_1,\kbar_2,\kbar_1',\kbar_2',M^2)~= ~
{{(\kbar^2_1+M^2)(
{\kbar^2_2}'+M^2)+(\kbar^2_2+M^2)(
{\kbar^2_1}'+M^2)}\over
{(\kbar_1-\kbar_1')^2+M^2}} ~~+ ~~\cdots
$$

We are interested in the infra-red limit in which the gluon mass 
$M \to 0$. We will, effectively, assume that two well-known leading-order 
properties of this limit generalize to all 
orders. The first property is that infra-red divergences,
due to the gluon particle poles in the reggeon states, interactions, and 
trajectory function, exponentiate to zero all diagrams that do not carry
zero color in the $t$-channel. The second property (which actually requires
appropriate behavior of the gauge coupling in higher orders) is that the
infra-red finitenes of color-zero reggeon interactions implies canonical 
scaling ($\sim Q^{-2}$) for color zero reggeon 
amplitudes in the limit that all internal
transverse momenta are scaled to zero. 

\section{Reggeon Diagrams for Helicity-Pole Limit Amplitudes}

For our purposes, ``maximal'' helicity-pole limits of multiparticle 
amplitudes are the most interesting to study. Because the Sommerfeld-Watson
representation involves only a single partial-wave amplitude, 
reggeon unitarity straightforwardly implies that reggeon
diagrams again appear. (Although~\cite{arw97} ``physical'' $k_{\perp}$ planes 
in general contain lightlike momenta !) As an example, 
we introduce variables for the 8-pt amplitude 
corresponding to the tree diagram of Fig.~3.
We consider the ``helicity-flip'' limit 

\noindent \parbox{2.8in}{\openup\jot
$$
 z,u_1,u^{-1}_2,u_3,u^{-1}_4 \to \infty
$$
(In this example, $u_1,u_2^{-1} \to \infty$ is a helicity-flip limit, while 
$u_1,u_2 \to \infty$ is a non-flip limit.) The behavior of invariants is
$$
\eqalign{&P_1.P_2 \sim u_1u^{-1}_2~, ~~~P_1.P_3 \sim u_1zu_3~, \cr
& P_2.P_4 \sim u^{-1}_2u^{-1}_4~, ~~~P_1.Q_3 \sim u_1z~, \cr
&Q_1.Q_3 \sim z~, ~~~P_4.Q_1 \sim zu^{-1}_4 
~ ~~ \cdots \cr
&~ P_1.Q,~P_2.Q,~ P_3.Q,~P_4.Q  ~~~\hbox{{\it finite }} }
$$}
\parbox{3.2in}{
\begin{center}
\leavevmode
\epsfxsize=2.7in
\epsffile{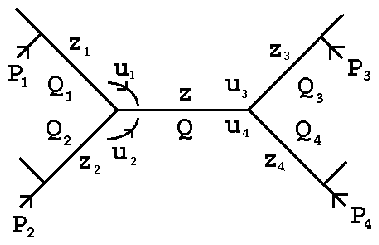}
\newline $~$
\newline Fig.~3 Variables for the 8-pt Amplitude
\end{center}}

\noindent \parbox{3in}{ \openup\jot 
Reggeon unitarity determines that the helicity-flip
limit is described by reggeon diagrams of the form shown in Fig.~4.
$~{\raisebox{-2mm}{\epsfxsize=0.3in \epsffile{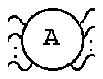}}}~$
contains all elastic scattering 
reggeon diagrams. The $T^F$ are new ``reggeon helicity-flip'' vertices
that play a crucial role in our QCD analysis. (These vertices do not appear 
in elastic scatttering reggeon diagrams).}
\parbox{3in}{
\begin{center}
\leavevmode
\epsfxsize=2.8in
\epsffile{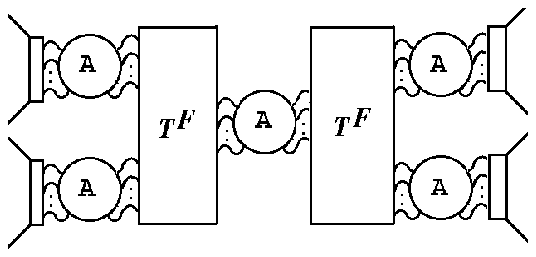}

Fig.~4 Reggeon Diagrams for the 8-pt 
\newline Amplitude
\end{center}
}

\section{Reggeon Helicity-Flip Vertices }

The $T^F$ vertices are most simply isolated kinematically by considering
a ``non-planar'' triple-regge limit which, for simplicity, we will define by 
introducing three distinct light-cone momenta. (This limit 
actually gives a sum of 
three $T^F$ vertices of the kind discussed above~\cite{arw97}, but in this 
talk we will not elaborate this subtlety.) We use the tree diagram of
Fig.~5(a) to define momenta and study the special kinematics 
\newline \parbox{2.5in}{
$$
\eqalign{&P_1\to (p_1,p_1,0,0),~~~p_1 \to \infty \cr
&P_2\to (p_2,0,p_2,0),~~~p_2 \to \infty \cr
&P_3\to(p_3,0,0,p_3),~~~p_3 \to \infty  \cr
&~ \cr
&Q_1\to (0,0,q_2, -q_3) \cr
&Q_2\to (0,-q_1,0,q_3) \cr
&Q_3\to (0,q_1,-q_2,0) }
$$
}
\parbox{3.5in}{
\begin{center}
\leavevmode
\epsfxsize=2.7in
\epsffile{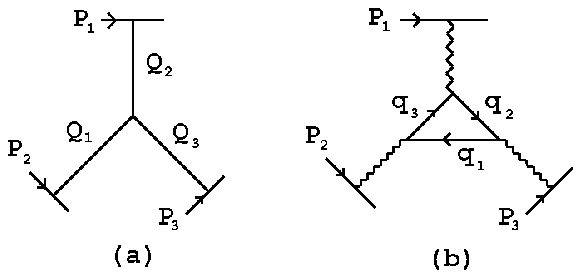}
\newline Fig.~5 (a) A Tree Diagram and (b) a quark loop coupling for three 
quark scattering.
\end{center}
}

We consider three quarks scattering
via gluon exchange with a quark loop coupling as in Fig.~5(b). The 
non-planar triple-regge limit 
$$
\to
~ g^6~~ { p_1p_2p_3 \over t_1 t_2 t_3 } ~~\Gamma_{1^+2^+3^+}(q_1,q_2,q_3)
~~\leftrightarrow ~~ g^3~~ { p_1p_2p_3 \over t_1 t_2 t_3 }~ T^F(Q_1,Q_2,Q_3) 
$$ 
where $~ \gamma_{i^+} = \gamma_0 + \gamma_i $ and 
$\Gamma_{\mu_1 \mu_2 \mu_3}$ is given by the quark triangle diagram i.e. 
$$
\Gamma_{\mu_1 \mu_2 \mu_3} = i\int {  d^4 k~ Tr \{ \gamma_{\mu_1}
(\st{q}_3 + \st{k} + m ) \gamma_{\mu_2} (\st{q_1} + \st{k} + m ) 
\gamma_{\mu_3} (\st{q}_2 + \st{k} + m) \} 
\over [ (q_1 + k)^2 - m^2 ][ (q_2 + k)^2 - m^2 ]
[ (q_3 + k)^2 - m^2 ]}
$$
where $m$ is the quark mass. We denote the $O(m^2)$ chirality-violating part of 
$~T^F ~(\equiv ~g^3 ~\Gamma_{1^+2^+3^+}~)$ by 
$T^{F,m^2}~$ and note that the 
limits $q_1, q_2, q_3 \sim Q \to 0$ and $m \to 0$ do not commute, i.e. 
$$
T^{F,m^2} {\centerunder{$\sim$}{\raisebox{-5mm} 
{$Q \to 0$} }}~Q ~i~m^2 \int {d^4k \over [ k^2 - m^2 ]^3 }
 ~~~~= ~  R ~Q 
$$
where $R$ is independent of $m$. This non-commutativity is an ``infra-red 
anomaly'' due to the triangle Landau singularity~\cite{cg}.

$T^F$ is also ultra-violet divergent. It is one of 
a general set of quark loop reggeon interactions that have ultra-violet 
divergences and so 
require regularization. We do this by introducing Pauli-Villars fermions 
that maintain the reggeon Ward identities that ensure gauge
invariance~\cite{arw97}. (Note that we take the regulator mass $m_{\Lambda}
\to \infty$ after $m \to 0$. This implies that the initial 
theory with $m \neq 0$ is non-unitary for $k_{\perp} \gsim m_{\Lambda}$.) 
For the regulated vertex, 
$T^{{\cal F},m^2}$, we obtain (for $m \neq 0$) 
$$
T^{{\cal F},m^2}(Q)~\sim ~ 
T^{F,m^2} ~- ~T^{F,m_{\Lambda}^2}
~~~{\centerunder{$\sim$}{\raisebox{-6mm} 
{$Q \to 0$} }} ~~~ Q^2
$$
However, since $T^{F,0} = 0$, we also have 
$$
T^{{\cal F},0}(Q) ~ ~~{\centerunder{$\sim$}{\raisebox{-6mm} 
{$Q \to 0$} }} ~~ -R ~Q
$$
showing that imposing reggeon Ward identities for $m \neq 0$ leads to a 
slower vanishing as $Q \to 0$ when $m=0$.

After color factors are included and all diagrams summed, we find that
$T^{{\cal F},0}(Q)$ survives only in very special reggeon vertices, i.e.
vertices coupling reggeon states with ``anomalous color parity''. We define 
color parity ($C_c$) via the transformation
$$
A^i_{ab} \to - A^i_{ba}
$$ 
for gluon color matrices. We say that a reggeon state has anomalous color 
parity if the signature ($\tau$), determined by whether the number of 
reggeons is even or odd, is not equal to the color parity. (Note 
that the 
reggeized gluon and the BFKL two reggeon state both have normal color 
parity.) 

We will be particularly interested in the ``anomalous odderon''
three-reggeon state with color factor
$f_{ijk}A^iA^jA^k$ that has $\tau = -1$ but $C_c = +1$ 
(c.f. the winding-number 
current
$K_{\mu}=\epsilon_{\mu \nu \gamma 
\delta}f_{ijk}A^i_{\nu}A^j_{\gamma}A^k_{\delta}$ ). $T^{{\cal F},0}(Q)$ 
\newline \parbox{2.3in}{ \openup\jot 
appears in the triple coupling of three anomalous 
odderon states as in Fig.~6.
The quantum numbers of the anomalous odderon 
state imply that, in this case, the survival of 
$O(m^2)$ processes as $m \to 0$ could be reproduced by the chirality 
violation of instanton interactions. In }
\parbox{4in}{
\begin{center}
\leavevmode
\epsfxsize=2.8in
\epsffile{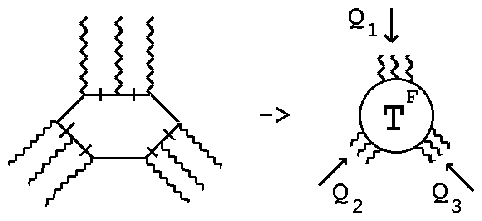}
\newline Fig.~6 An Anomalous Odderon 
\newline Triple Coupling.
\end{center}
}
\newline our case the presence of such non-perturbative interactions in the 
massless theory is due to our regularization procedure for reggeon 
interaction vertices.
 
\section{A Quark Mass Infra-Red Divergence} 

A vital consequence of the ``anomalous'' behavior of $~T^{{\cal F},0}~$ 
as $~Q \to 0~$ is that an 
additional infra-red divergence 
is produced (as $m \to 0$) in massless gluon reggeon diagrams. 
The divergence occurs in diagrams involving the $T^F$ 
where $~Q_1 \sim Q_2 \sim Q_3 \sim 0
~$ is part of the integration region. However, the $T^F$ only appear in 
vertices coupling distinct reggeon 
\newline \parbox{2.5in}{ channels. 
A potentially divergent diagram containing $T^F$
vertices is shown in
Fig.~7. In this diagram an anomalous odderon reggeon state is denoted by
$~\raisebox{-1.5mm}{$\epsfxsize=1.3in 
\epsffile{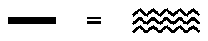}$}~ $ while 
$~\raisebox{-0.5mm}{$\epsfxsize=0.3in \epsffile{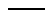}$}~$
denotes any normal reggeon state. Fig.~7 is of the general form illustrated 
in Fig.~4, except that we are allowing the vertices $V_i$ to involve more 
complicated external states than a single scattering quark. 
The canonical 
scaling of the anomalous odderon states 
gives the infra-red behaviour 
}
\parbox{3.5in}{ 
\begin{center}
\leavevmode
\epsfxsize=2.9in 
\epsffile{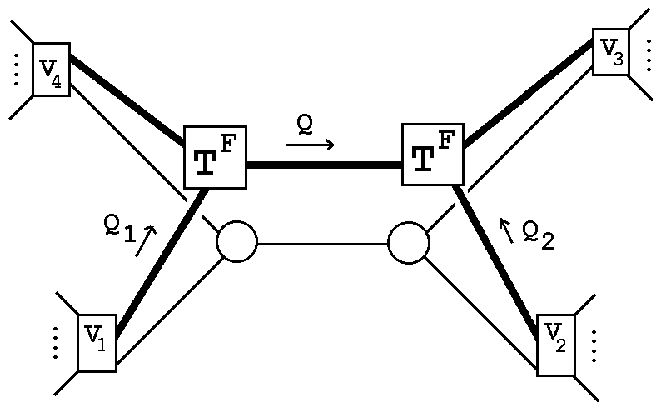}
\newline Fig.~7 A Divergent Reggeon Diagram 
\end{center}
}
$$
\eqalign{& \int \cdots {d^2Q_1 \over Q_1^2}~{d^2Q_2 \over Q_2^2} 
~{d^2Q \over Q^2 (Q-Q_1)^2 (Q - Q_2)^2 }\cr
&\times  V_1(Q_1)V_2(Q_2) V_3(Q- Q_2) V_4(Q - Q_1)~ T^{ {\cal F}}(Q_1,Q)
T^{ {\cal F}}(Q,Q_2)\cr
&\times~~\hbox{[regular vertices and reggeon propagators]}
}
$$
Depending on the behaviour of the $V_i~$ it appears that a
divergence potentially occurs when $Q \sim Q_1 \sim Q_2 \to 0$.
However, in general gauge invariance produces~\cite{arw97}
a cancelation of this 
divergence by a similar divergence of diagrams related to that of 
Fig.~7 by reggeon Ward identities for the reggeons within the anomalous odderon 
states. 

We can preserve the divergence of the diagram of Fig.~7 and 
eliminate the possibility of a cancelation 
if we partially break the SU(3) gauge symmetry to SU(2). In effect, introducing 
the symmetry breaking mass scale provides a scale for the logarithmic quark 
mass divergence. (Note also that the topology of an instanton is defined
with respect to an SU(2) subgroup. As a result we anticipate that this
partial breaking enhances the significance of topological effects associated
with the anomaly.) We can show that other diagrams can not cancel the quark
mass divergence discussed above if
$~\raisebox{-0.5mm}{$\epsfxsize=0.3in \epsffile{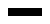}$}~$
is any SU(2) singlet 
combination of massless gluons with 
$~C_c= -\tau = +1~ $ (i.e. a generalized SU(2) anomalous odderon) and  
$~\raisebox{-0.5mm}{$\epsfxsize=0.3in \epsffile{dss16.ps}$}~$
is a normal reggeon state 
containing one or more SU(2) singlet 
massive reggeized gluons (or quarks). We can then regard Fig.~7 as containing 
only reggeon states of the form 
$~\raisebox{-2mm}{$\epsfxsize=0.4in \epsffile{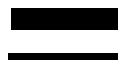}$}~$
coupling via a combination of regular and $~T^{{\cal F},0}~$ vertices.

We must also discuss specifically the behavior of the $V_i$.
A-priori reggeon Ward identities imply $ V_i \to 0$ when 
$Q_i \to 0$. This, in itself, would be sufficient to eliminate any
divergence in Fig.~7 ! However, if we impose the ``initial condition'' that 
$V_1,V_2 ~ \st{\rightarrow}~ 0$, the divergence is present and in fact 
occurs similarly in a general class of diagrams, as we now discuss. 
We 
\newpage
\noindent consider a diagram having the general structure 
illustrated in Fig.~8, in which there are 
$n + 3$ multi-reggeon states of the form
$~\raisebox{-2mm}{$\epsfxsize=0.4in \epsffile{cspp12.ps}$}~$.
Imposing $V_1,V_2 ~ \st{\rightarrow}~ 0$ and assuming that 
\newline \parbox{2in}{
reggeon Ward identities are satisfied by the remaining vertices, i.e. 
$$
V_i(Q_i) \sim V(Q_i) = Q_i 
$$
$i \neq 1,2$, gives that 
Fig.~8 
has the infra-red behavior
$$ 
\int {d^2 Q \over Q^2}~\left[\int {d^2 Q \over Q^4}\right]^n 
$$
$$
\times \left[V(Q)~ T^ {\cal F}(Q)\right]^n 
$$
Thus giving (as $m \to 0$) an 
}
\parbox{4in}{
\begin{center}
\leavevmode
\epsfxsize=3.6in
\epsffile{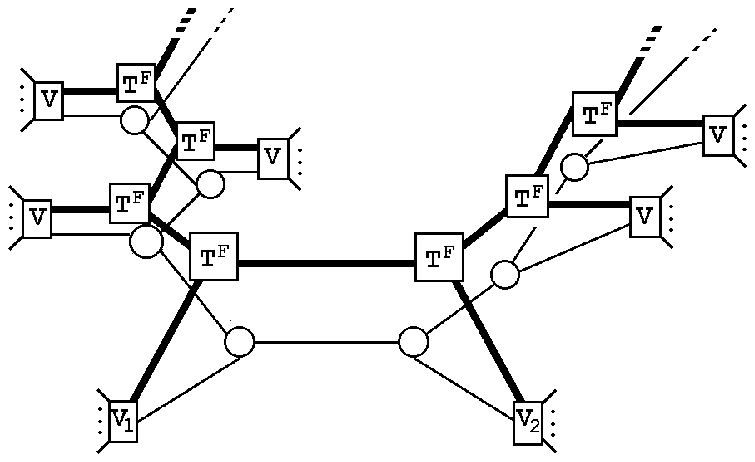}

Fig.~8 A General Divergent Diagram
\end{center}
}
\newline overall logarithmic divergence. 
In general, this divergence occurs in just 
those multi-reggeon 
diagrams which contain only SU(2) color zero
states of the form 
$~\raisebox{-2mm}{$\epsfxsize=0.4in \epsffile{cspp12.ps}$}~$
coupled by regular and $~T^{{\cal F},0}~$ vertices, as in the examples we 
have discussed.

\section{Confinement and a Parton Picture}

We define physical amplitudes from divergent diagrams by extracting the 
coefficient of the logarithmic divergence. The resulting reggeon states and 
amplitudes produce ``a confinement phenomenon'' in the following sense.
A particular set of color-zero reggeon states is selected that contains no
massless multigluon states and has the necessary completeness property to
consistently define an S-Matrix. By completeness we mean that if two or more
of this set of states initially scatter 
via QCD interactions, the final states contain only 
\newline \parbox{2.2in} {
arbitrary numbers of the
same set of states. Since the divergence involves zero 
$k_{\perp}$ for the anomalous odderon component of
each reggeon state, an ``anomalous odderon condensate'' 
is generated. The general picture
is illustrated in Fig.~9. 
In addition to the zero $k_{\perp}$ (or wee-parton) component,
physical reggeon states have a 
finite momentum ``normal'' parton component carrying 
}
\parbox{3.8in}{
\begin{center}
\leavevmode
\epsfxsize=3.6in
\epsffile{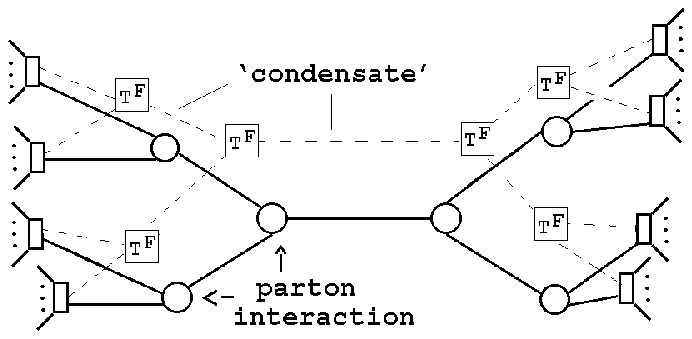}
\newline Fig.~9 The Parton Picture
\end{center}
}
the kinematic properties of interactions. 
We emphasize that the ``scattering'' of the $k_{\perp} = 0$ condensate is
directly due to the infra-red quark triangle anomaly. 

The breaking of the gauge symmetry has produced physical states in which 
the ``partons'' are separated into a universal wee-parton component and a 
normal reggeon parton component which is distinct in each distinct 
physical state. However, the condensate has the important 
property that it switches the signature compared to that of the normal parton 
component. We note the following important reggeon states. 

\begin{itemize}

\item{There is a ``pomeron'' whose normal parton component is a 
reggeized gluon. This is a Regge pole with $\tau  = - C_c = + 1$ 
and intercept $\neq 0$. }

\item{ A bound-state reggeon formed from two massive SU(2) doublet gluons
gives an exchange-degenerate partner
to the pomeron, i.e. a Regge pole with $\tau  = - C_c = - 1$ . The SU(2) 
singlet massive gluon lies on this trajectory. }

\item{ ``Hadrons'' have a constituent quark normal parton component.}

\end{itemize}
All the features of supercritical RFT are present, but we will not discuss 
the details. Although we have not specifically discussed quark reggeons, 
the hadron reggeons they form are vital. ``Stability'' of the quark parton
component within the condensate produces chiral symmetry breaking, which
determines that hadrons are not eigenstates of color parity. This is
necessary if the exchange of a pomeron with $C_c = -1$ is to
describe elastic scattering. 

\section{Restoration of SU(3) Gauge Symmetry }

We make only a few brief comments on this, obviously important, subject.
Because of complimentarity~\cite{fs}, restoring SU(3)
symmetry (which involves decoupling a color triplet Higgs
scalar field) should 
be straightforward if we impose a transverse momentum cut-off 
$k_{\perp} < \Lambda_{\perp}$. Restoring the symmetry involves
removing the 
mass scale that distinguishes normal (finite momentum) partons from wee 
(zero momentum) partons and produces 
the reggeon condensate. Mapping the (partially) broken theory completely
onto supercritical RFT implies that the condensate 
and the odd-signature partner for the pomeron disappear simultaneously. The 
result is then the critical pomeron~\cite{cri}. The wee-parton condensate
will be replaced by a small $k_{\perp}$, wee parton, 
critical phenomenon that merges smoothly with the large 
$k_{\perp}$ normal (or constituent) parton component of physical states,
just as originally envisaged by Feynman~\cite{rf}. (Note that, because of
the odd SU(3) color charge parity of the pomeron, the two-gluon BFKL pomeron 
will not be involved.) 

Mapping partially-broken QCD onto supercritical RFT has further 
consequences. In particular, it implies that the 
$\Lambda_{\perp}$ scale mixes with the symmetry breaking scale 
and becomes a ``relevant parameter'' for the
critical behavior. It then follows that, after the symmetry breaking scale is 
removed, there will (for a general number of quark flavors) be a 
$~\Lambda_{\perp c}$ such that 
$\Lambda_{\perp} > \Lambda_{\perp c} ~$ implies the pomeron is in the 
subcritical phase, while $\Lambda_{\perp} <  \Lambda_{\perp c}~ $ implies it 
is in the 
supercritical phase. 
We conclude that the supercritical phase can be
realized with the gauge symmetry restored if $\Lambda_{\perp} $ is taken 
small enough. However, $\alpha_{\spom}(0)$ and the mass of the exchange
degenerate, composite, ``reggeized gluon'' will be functions of
$\Lambda_{\perp}$. In deep-inelastic diffraction, large $Q^2 $ will act as a
(local) lower $k_{\perp}$ cut-off and produce a ``finite volume'' effect
that can keep the theory supercritical as the SU(3) symmetry is
restored. 

To remove $\Lambda_{\perp}$ requires $\Lambda_{\perp c} ~= \infty $. We will 
not discuss here why we believe this requires a specific quark flavor
content. It is interesting that, for any quark content, we can take 
$\Lambda_{\perp} <<  \Lambda_{\perp c} ~  $, and go deep into the 
supercritical phase. We obtain a picture in which constituent quark 
hadrons interact via a massive composite ``gluon'' (and an exchange 
degenerate pomeron). Confinement and chiral symmetry breaking are realized 
via a simple, universal, wee parton component of physical states. This is 
remarkably close to the realization of the constituent quark model
via light-cone quantization that has been advocated by light-cone 
enthusiasts~\cite{kw}. 

\vspace{0.2in}

\noindent { \bf References}


\begin{thebibliography}{99}

\bibitem{h1} H1 Collaboration, pa02-61 ICHEP'96 (1996), For a final version
of the analysis see {\it Z. Phys.} {\bf C76}, 613 (1997).

\bibitem{arw80} A.~R.~White, CERN preprint TH.2976 (1980). A summary of this 
paper is presented in the Proceedings of the
XVIth Rencontre de Moriond, Vol.~2 (1981). 

\bibitem{arw97} A.~R.~White, hep-ph/9712466 (1997).

\bibitem{kw} K.~G.~Wilson, T.~S.~Walhout, A.~Harindranath, Wei-Min Zhang, 
S.~D.~Glazek and R.~J.~Perry, {\it Phys. Rev.} {\bf D 49}, 6720 (1994);
L. Susskind and M. Burkardt, hep-ph/9410313 (1994).

\bibitem{gpt} V.~N.~Gribov, I.~Ya.~Pomeranchuk and K.~A.~Ter-Martirosyan,
{\it Phys. Rev.} {\bf 139B}, 184 (1965).

\bibitem{arw1} A.~R.~White, Int. J. Mod. Phys. {\bf A11}, 1859 (1991);
 A.~R.~White in {\em Structural Analysis of Collision Amplitudes},
(North Holland, 1976). 

\bibitem{bfkl} E.~A.~Kuraev, L.~N.~Lipatov, V.~S.~Fadin, {\it Sov. Phys.
JETP} {\bf 45}, 199 (1977) ; Ya.~Ya.~Balitsky and L.~N.~Lipatov, {\it Sov. J.
Nucl. Phys.} {\bf 28}, 822 (1978). 
V.~S.~Fadin and L.~N.~Lipatov, {\it Nucl. Phys.} {\bf B477},
767 (1996) and further references therein.

\bibitem{bs} J.~B.~Bronzan and R.~L.~Sugar, {\it Phys. Rev.} {\bf D17}, 
585 (1978). This paper organizes into reggeon diagrams the results from 
H.~Cheng and C.~Y.~Lo, {\it Phys. Rev.} {\bf D13}, 1131 (1976), 
{\bf D15}, 2959 (1977). 

\bibitem{fs1} V.~S.~Fadin and V.~E.~Sherman, {\it Sov. Phys.} {\bf  JETP 45}, 
861 (1978).

\bibitem{cg} S.~Coleman and B.~Grossman, {\it Nucl. Phys. }
{\bf B203}, 205 (1982).

\bibitem{cri}  A.~A.~Migdal, A.~M.~Polyakov and K.~A.~Ter-Martirosyan, 
{\it Zh. Eksp. Teor.  Fiz.} {\bf 67}, 84 (1974); 
H.~D.~I.~Abarbanel and J.~B.~Bronzan, {\it Phys. Rev.} {\bf D9}, 2397 (1974).

\bibitem{sw} H.~P.~Stapp 
in {\em Structural Analysis of Collision Amplitudes},
(North Holland, 1976); 
H.~P.~Stapp and A.~R.~White, {\it Phys. Rev.} {\bf D26}, 2145 
(1982). 

\bibitem{fs} E.~Fradkin and S.~H.~ Shenker, {\it Phys. Rev. }
{\bf D19}, 3682 (1979); 
T.~Banks and E.~Rabinovici, {\it Nucl. Phys. } {\bf B160}, 349 (1979).
 
\bibitem{rf} R.~P.~Feynman in {\it Photon Hadron Interactions}
(Benjamin, 1972)).

\end{thebibliography}
\end{document}